\documentclass[12pt]{article} 

\usepackage{geometry} 
\usepackage{graphicx} 
\usepackage{caption}
\usepackage{subcaption}

\usepackage{booktabs} 
\usepackage{array} 
\usepackage{paralist} 
\usepackage{verbatim} 
\usepackage{subfig} 

\usepackage{fancyhdr} 
\usepackage{sectsty}
\allsectionsfont{\sffamily\mdseries\upshape} 

\usepackage[nottoc,notlof,notlot]{tocbibind} 
\usepackage[titles,subfigure]{tocloft} 

\usepackage{amsfonts}

\usepackage{amsmath}
\DeclareMathOperator{\Tr}{Tr}

\title{The Emergence of Supersymmetry in $\gamma_i$-deformed ${\cal N}=4$ super-Yang-Mills theory}

\author{Qingjun Jin\footnote{qxj103@psu.edu}\\[6pt]
Institute for Gravitation and the Cosmos\\
 The Pennsylvania State University \\
University Park PA 16802, USA \\ }

\date{} 

\begin{document}
\maketitle

\begin{abstract}

The $\gamma_i$-deformed $\mathcal{N}=4$ super-Yang-Mills theory is a non-supersymmetric  deformation of the maximally-supersymmetric gauge theory in four dimensions which is conformally-invariant at the planar level. At the non-planar level non-trivial renormalization of marginal double-trace operators appears to break conformal invariance.
We study the one loop renormalization of the theory and the resulting flow of Yukawa and scalar coupling at the order 
${\cal O}(\frac{\lambda}{N^2})$, and find $\gamma_i$ parameters receive nontrivial quantum corrections. We show that, at low energies,  
the $\gamma_i$ parameters flow to a common value and $\gamma_i$-deformed theory reaches an $\mathcal{N}=1$-supersymmetric fixed point
described by the $\beta$-deformed $\mathcal{N}=4$ super Yang-Mills theory.

\end{abstract}

\newpage

\section{Introduction}

Shortly after the discovery of the correspondence between string theory on AdS$_5\times$S$^5$ and ${\cal N}=4$ super-Yang-Mills (sYM)
theory \cite{Maldacena:1997re, Gubser:1998bc, Witten:1998qj} it was realized that it can be used to identify four-dimensional gauge theories with 
reduced or no supersymmetry which are conformally invariant in the planar limit. To this end the geometric S$^5$ factor is replaced with 
a less symmetric space M$^5$ while adjusting the other fields such that AdS$_5\times$M$^5$ is a solution of the supergravity equations 
of motion.
For example, by choosing M$^5=$S$^5/\Gamma$ with $\Gamma\subset \text{SU}(n)\subset\text{SU}(4)$ in a regular representation it follows that 
\cite{Kachru:1998ys, Lawrence:1998ja} ${\cal N}=4-n$ should be conformally invariant in the planar limit. Evidence for this was provided 
in \cite{Bershadsky:1998cb, Bershadsky:1998mb} to all orders in planar perturbation theory.
Another example is the Lunin-Maldacena background \cite{Lunin:2005jy} and its non-supersymmetric generalization \cite{Frolov:2005dj} 
which describe, respectively, the $\beta$-deformed ${\cal N}=4$ sYM theory \cite{Leigh:1995ep}  and a three-parameter generalization 
known as the $\gamma_i$-deformation.

While different in details, the $\gamma_i$-deformation and the regular non-supersymmetric orbifolds of ${\cal N}=4$ 
sYM theory share most if not all of their planar properties. For example, their planar scattering amplitudes are inherited 
up to trivial overall phase or numerical factors from the parent theory; moreover, their planar dilatation operators define 
integrable Hamiltonians \cite{Beisert:2005if, Beisert:2005he}. Strong coupling integrability of the former was discussed in 
detail in \cite{Frolov:2005dj} while that of the latter follows straightforwardly from~\cite{Bena:2003wd}.

Explicit calculations for particular \cite{Tseytlin:1999ii, Csaki:1999uy} and general \cite{Dymarsky:2005uh} non-supersymmetric orbifolds 
of ${\cal N}=4$ sYM theory showed that dimension-four double-trace operators acquire a nonzero $\beta$-function in all such theories 
thus breaking conformal invariance. It was also shown in \cite{AD_RR_up, Fokken:2013aea} that a similar phenomenon occurs in 
the $\gamma_i$-deformed ${\cal N}=4$ sYM theory. In both  the one-loop beta function has no zeroes; together with the vanishing beta function
for the gauge coupling at ${\cal O}(1/N)$ this renders the theory unstable. An interpretation of this instability for the $\gamma_i$-deformed theory 
was suggested in \cite{Fokken:2013aea}; for orbifolds this was discussed in \cite{AD_SF_IK_RR_up} where it was suggested that the theory 
flows to a fixed point described by an ${\cal N}=4$ sYM theory with a gauge group of reduced rank. Strong coupling evidence in this direction 
was given in \cite{Horowitz:2007pr}.

Since it depends on more parameters, the flow of the  $\gamma_i$-deformed ${\cal N}=4$ sYM theory can in principle be different 
from that of the orbifold theories. In this paper we shall explore this flow. By analyzing the non-planar one-loop four-point scattering 
amplitudes with various external states we will show that in the absence of supersymmetry the deformation parameters $\gamma_i$ 
are renormalized nontrivially. If the ensuing flow starts with $\gamma_i$ differing by ${\cal O}(\frac{1}{N^2})$, the theory has a fixed point 
described by the ${\cal N}=1$ $\beta$-deformed $\mathcal{N}=4$ sYM theory with $\beta$ parameter being the average of the initial 
three $\gamma_i$ parameters. The logarithmic running of double-trace operators found in \cite{Dymarsky:2005uh, Fokken:2013aea} is 
also suppressed by the  flow of $\gamma_i$ parameters.

We begin in section \ref{section:gamma} with a brief introduction of $\gamma_i$-deformed sYM theory. In section 
\ref{section:betafromdivergence} we describe a method to extract beta functions directly from on-shell scattering 
amplitudes. Using this method in sections \ref{section:yukawa} and \ref{section:scalar} we discuss the flow of Yukawa and 
scalar couplings, respectively. We close in section \ref{section:discussion} with a summary of our results and further remarks.


\section{The $\gamma_i$-deformed super-Yang-Mills theory  \label{section:gamma}}

As discussed in \cite{Lunin:2005jy}, the $\beta$-deformed $\mathcal{N}=4$ sYM theory can be interpreted as a non-commutative deformation
of the maximally-supersymmetric YM theory in the R-symmetry directions; the corresponding star-product of two fields $\Phi_1$ and $\Phi_2$
is given by 
\begin{eqnarray}
\Phi_1*\Phi_2 = e^{i\beta(Q^1(\Phi_1)Q^2(\Phi_2)+Q^2(\Phi_1)Q^3(\Phi_2)+Q^3(\Phi_1)Q^1(\Phi_2))}\Phi_1\Phi_2 \ .
\end{eqnarray}
Here $\Phi$ stands for any of the fields of ${\cal N}=4$ theory and $Q^i(\Phi)$ are the charges of the field $\Phi$ under the three 
Cartan generators of the SU$(4)$ R-symmetry group:
\begin{equation}
\begin {tabular} {|c|c|c|c|c|c|c|c|c| }
\hline
&A&$\psi^1$&$\psi^2$&$\psi^3$&$\psi^4$&$\phi^{14}$&$\phi^{24}$&$\phi^{34}$\\
\hline
$Q^1$&0&$\frac{1}{2}$&$-\frac{1}{2}$&$-\frac{1}{2}$&$\frac{1}{2}$&1&0&0\\
\hline
$Q^2$&0&$-\frac{1}{2}$&$\frac{1}{2}$&$-\frac{1}{2}$&$\frac{1}{2}$&0&1&0\\
\hline
$Q^3$&0&$-\frac{1}{2}$&$-\frac{1}{2}$&$\frac{1}{2}$&$\frac{1}{2}$&0&0&1\\
\hline
\end {tabular}\label{qs}
\end{equation}
The construction of its string theory dual \cite{Lunin:2005jy} suggested that a natural generalization \cite{Frolov:2005dj} is to introduce three 
deformation parameters -- $(\gamma^1, \gamma^2, \gamma^3)$, one for each  pair of charges -- while keeping the interpretation as a 
non-commutative deformation. The corresponding star-product is then
\begin{eqnarray}
\label{starproduct_gamma}
&&
\Phi_1*\Phi_2 = e^{i \theta(\Phi_1,\Phi_2)}\Phi_1\Phi_2
\qquad
[\Phi_1,\Phi_2]_\theta = e^{i\theta(\Phi_1\Phi_2)}\Phi_1\Phi_2-e^{-i\theta(\Phi_1\Phi_2)}\Phi_2\Phi_1
\\
&&\theta(\Phi_1\Phi_2)=\epsilon_{lmn}Q^l(\Phi_1)Q^m(\Phi_2)\gamma^n \ .
\end{eqnarray}
R-charge conservation implies that the phase deforming the product of $n$ fields, $\Phi_1,\dots, \Phi_n$, is
\begin{equation}
\theta(\Phi_1\Phi_2\cdots\Phi_n)=\sum_{i<j}\theta(\Phi_i\Phi_j) \ .
\end{equation}
For later convenience we also define the deformed color trace,
\begin{equation}
\Tr_{\theta}(\Phi_1\Phi_2\cdots\Phi_n)\equiv e^{i\theta(\Phi_1\Phi_2\cdots\Phi_n)}\Tr(\Phi_1\Phi_2\cdots\Phi_n) \ .
\end{equation}

The single-trace part of the Lagrangian can be obtained by replacing commutators in $\mathcal{N}=4$ sYM theory by deformed 
commutators \eqref{starproduct_gamma}:
\begin{equation}
\begin{aligned}
L_{\gamma}^{planar}=\Tr\Bigl(&-\frac{1}{4}F_{\mu\nu}F^{\mu\nu}-\frac{1}{4}D_{\mu}\bar{\phi}_{AB} D^{\mu}\phi^{AB}
+\frac{g^2}{16}[\bar{\phi}_{AB},\ \bar{\phi}_{CD}]_{\theta}[\phi^{AB},\ \phi^{CD}]_{\theta}\\
&+i\bar{\psi}_A\bar{\sigma}^{\mu}D_{\mu}\psi^A+\frac{ig}{\sqrt{2}}\left(\bar{\psi}_A\left[\phi^{AB},\ \bar{\psi}_B\right]_{\theta}
+\psi^A\left[\bar{\phi}_{AB},\ \psi^B\right]_{\theta}\right)\Bigr).\\
\end{aligned}\label{gammaplanarl}
\end{equation}
This is however not the complete Lagrangian and additional double-trace terms must be supplied separately. To see that this is 
the case it suffices to inspect the supersymmetric limit, $\gamma_1=\gamma_2=\gamma_3=\beta$ \cite{Leigh:1995ep}. In this case
the Lagrangian is
\begin{equation}
\begin{aligned}
L_\beta&=\Tr\left(-\frac{1}{4}F_{\mu\nu}F^{\mu\nu}-D_{\mu}\bar{\phi}_{i} D^{\mu}\phi^{i}
+i\bar{\psi}_i\bar{\sigma}^{\mu}D_{\mu}\psi^i+i\bar{\psi}_4\bar{\sigma}^{\mu}D_{\mu}\psi^4-\frac{g^2}{2}[\phi^{i},\ \bar{\phi}_{i}]^2\right.\\
&\hspace{0mm}-\left.\frac{ih}{\sqrt{2}}\left(\epsilon^{ijk}\bar{\psi}_i\left[\bar{\phi}_j,\ \bar{\psi}_k\right]_{\theta}+\epsilon_{ijk}\psi^i[\phi^j,\ \psi^k]_{\theta}\right)
+\sqrt{2}ig\left(\bar{\psi}_i\left[\phi^{i},\ \bar{\psi}_4\right]+\psi^i\left[\bar{\phi}_{i},\ \psi^4\right]\right)\right)\\
&+h^2\Tr\left([\phi^{i},\ \phi^{j}]_{\theta}T^a\right)
\Tr\left([\bar{\phi}_{i},\ \bar{\phi}_{j}]_{\theta}T^a\right),\\
\end{aligned}\label{lbeta}
\end{equation}
The double-trace terms arise from the integration over the U$(1)$ part of the $F_i$-auxiliary fields  which does not decouple in 
the presence of a non-trivial star product; the renormalization of the Lagrangian \eqref{gammaplanarl} implies that similar terms
must be added there as well \cite{AD_RR_up, Fokken:2013aea}.
The $\beta$-deformed sYM theory is a conformal field theory even at non-planar level if the parameters $h$ and $\beta$ and the gauge 
coupling $g$ are not independent \cite{Leigh:1995ep}.  Through two-loop level  this relation is \cite{Freedman:2005cg}
\begin{equation}\label{h0}
\frac{g^2}{h^2} = 1-\frac{4\sin^2\beta}{N^2} \ .
\end{equation}
A relation with similar consequences does not seem to exist for the non-supersymmetric $\gamma$-deformed theory. We will argue 
in later sections that the theory nevertheless flows to a non-trivial infrared fixed point.

\section{From Divergence of Amplitudes to Beta Function\label{section:betafromdivergence}}

The textbook approach to $\beta$-function calculations makes use of the renormalization factors $Z$. For example, in pure Yang-Mills 
theory, one computes gauge field and coupling renormalization factors $Z_A$ and $Z_g$ and extracts the $\beta$-function 
for the gauge coupling form the combination $Z_A^{-3}Z_g^2\mu^\epsilon$ relating the bare and renormalized couplings.\footnote{It is, of 
course, technically easier to use the ghost vertex and the ghost wave function renormalization factors to avoid the evaluation of the one-loop 
correction to the three-gluon vertex.}
On-shell methods can be used to find on-shell counterterm matrix elements which may then be reinterpreted in terms of
Lagrangian counterterms and renormalization factors and thus may be used to extract beta functions.
In this section we describe an alternative way which allows us to extract $\beta$-functions directly from on-shell amplitudes.

Let us start with a  Lagrangian $L=L(\Phi_a, g_i)$ with general fields content $\Phi_a$ and coupling constants $g_i$. 
The bare and  renormalized Lagrangian are
\begin{equation}
\label{lz}
L_Z=L(\Phi^0_i,\ g_i^0)=L(Z_a^{\frac{1}{2}}\Phi_a, \ \tilde{Z}_ig_i\tilde{\mu}^{-d_i\epsilon}),
\end{equation}
where $d_i\epsilon$ is the dimension of the coupling $g_i$ in $4-\epsilon$ space-time dimension, 
$\tilde{Z}_i$ is a combination of $Z_i$'s and $Z_a$'s which can be read from comparing two expressions for $L_Z$ in \eqref{lz}. 
For example, for gauge coupling in Yang-Mills theory, $\tilde{Z}_g=Z_A^{-\frac{3}{2}}Z_g$.

A one-loop $n$-point amplitude $\mathcal{A}^{1-loop}(12\cdots n)$ has two components: the first is the one-loop amplitude 
computed with respect to the original Lagrangian and the second is the contribution of counterterms,
\begin{equation}
\mathcal{A}^{1-loop}=\mathcal{A}^{1-loop}(g_i)+\mathcal{A}_{ct} \ .
\label{genA1loop}
\end{equation}
The counterterm contribution can be written as the difference between the tree-level amplitudes of the Lagrangian $L_Z$ and the tree amplitudes
of the Lagrangian with $Z_i=1$. Since the purpose of the one-loop counterterms is to cancel the divergences of the one-loop amplitudes we can 
express $\mathcal{A}_{ct} $ in \eqref{genA1loop} as
\begin{equation}\label{cancel}
\mathcal{A}_{ct} = \mathcal{A}^{tree}(\tilde{Z}_ig_i)-\mathcal{A}^{tree}(g_i)=-\frac{1}{\epsilon}Div(g) \ ,
\end{equation}
where the right-hand side stands for the negative of the divergence of the corresponding one-loop amplitude which may depend on all the couplings of the theory generically denoted by argument ${\hat g}$.

At one loop, only graphs with a single counterterm insertion (or a single nontrivial term in $Z$ factors) are relevant; parametrizing the 
$Z_i$ factors as~\footnote{As in the case of $Div_i({\hat g})$, the argument of ${\cal D}_i$ emphasizes the fact that ${\cal D}_i$ can depend 
on all the couplings of the theory.}
\begin{equation}
\tilde{Z}_i=1+\frac{1}{\epsilon}\mathcal{D}_i({\hat g}) \ ,
\end{equation}
we can therefore expand eq.~\eqref{cancel} to linear order in ${\cal D}_i({\hat g})$:
\begin{equation}\label{cancelf}
\sum_i\frac{\partial \mathcal{A}_\alpha^{tree}(g_i)}{\partial g_i}\;g_i{\cal D}_i({\hat g})+Div_{\alpha}({\hat g})=0\ ,
\end{equation}
where the sum runs over all the parameters of the theory that receive nontrivial infinite renormalization and the index $\alpha$
runs over the set of all amplitudes of the theory.

Thus, by examining sufficiently many amplitudes $\mathcal{A}_\alpha$ and determining their divergent parts it should 
be possible to find the renormalization factors by solving the linear system made from the equations above. 

The $\beta$-function of the coupling $g_i$ is related to $\mathcal{D}_i({\hat g})$ in the usual way following from the relation between the 
bare and the renormalized couplings $g_i^0=\tilde{Z}_i\tilde{\mu}^{-d_i\epsilon}g_i$:
\begin{equation}
\begin{aligned}
&\frac{1}{\epsilon}\frac{d \mathcal{D}_i({\hat g})}{d\ln\mu}=d_i\epsilon -\frac{1}{g_i}\frac{d g_i}{d\ln \mu}=-\frac{1}{g_i}\beta(g_i)
~~\Rightarrow~~
\beta(g_i)=-\frac{g_i}{\epsilon}\frac{d \mathcal{D}_i({\hat g})}{d\ln\mu} \ ,
\end{aligned}\label{betagi}
\end{equation}

$\mathcal{D}_i({\hat g})$ is a polynomial of coupling constants,
\begin{equation}\label{Di}
\mathcal{D}_i({\hat g})=\sum_Ac_A\prod_j g_j^{n_{Aij}}\ ,\hspace{5mm}
\frac{d \mathcal{D}_i({\hat g})}{d\ln\mu}=\epsilon\sum_Ac_A\sum_kn_{Aik}d_k\prod_j g_j^{n_{Aij}}+\cdots\ ,
\end{equation}

To simplify \eqref{Di}, we need to find the classical scaling dimension\footnote{The scale dependence of $\mathcal{D}_i({\hat g})$ is introduced by that of $g_j$. The scale dependence of $g_i$ is $\frac{dg_i}{d\ln\mu}=\epsilon d_ig_i+\beta(g)$, and the first term gives the classical scaling dimension of $g_i$.} of $\mathcal{D}_i({\hat g})$. Start with the all loop scattering amplitude\footnote{In order to find classical scaling dimensions, Z factors are not needed in \eqref{allloop}.},
\begin{equation}\label{allloop}
\mathcal{A}(g_i\tilde{\mu}^{-d_i\epsilon})=\mathcal{A}^{tree}(g_i\tilde{\mu}^{-d_i\epsilon})+\mathcal{A}^{1-loop}(g_i\tilde{\mu}^{-d_i\epsilon})+\cdots\ .
\end{equation}
A one loop amplitude contains two components with non-zero classical scaling dimensions which the tree amplitude does not have. One is $\mathcal{D}_i(g_j\tilde{\mu}^{-d_j\epsilon})$, the other is a loop integral which scales as $\int\frac{d^{4-\epsilon}l}{(l^2)^2}\sim \mu^{-\epsilon}$. The classical scaling dimensions of these two components must cancel each other, since tree amplitude and loop amplitude have the same classical scaling dimensions:
\begin{equation}\label{nijdj}
\mathcal{D}_i(g_j\tilde{\mu}^{-d_j\epsilon})=\mathcal{D}_i(g_j)\tilde{\mu}^{\epsilon}
~~\Rightarrow~~
\sum_j n_{Aij}d_j=-1 \ ,
\end{equation}
Combining \eqref{betagi}, \eqref{Di} and \eqref{nijdj}, we find a simple relation between the beta function and the divergence,
\begin{equation}\label{betagi1}
\beta(g_i)=g_i\mathcal{D}_i.
\end{equation}
Using this relation, we can rewrite \eqref{cancelf} as\footnote{We dropped $\mathcal{O}(\epsilon)$ terms in \eqref{betadiv}.},
\begin{equation}\label{betadiv}
\frac{d}{d\ln\mu}\mathcal{A}^{tree}_{\alpha}(g_i)=-Div_{\alpha}(g_i).
\end{equation}

Eq.~\eqref{betadiv} provides a direct relation between one-loop divergence of an amplitude and and $\beta$-functions of coupling constants. 
If there are in all $N_g$ independent coupling constants, in order to solve all $\beta(g_i)$'s from \eqref{betadiv}, one need to find at least $N_g$ 
independent equations. In practice, a proper choice of equations will greatly simplify the calculation. For example, in $\gamma_i$-deformed SYM 
practically all $\mathcal{A}^{tree}_{\alpha}$'s will depend on $g$, the gauge coupling. But the four-gluon amplitude does not depend on any other 
coupling constants, so one can solve $\beta(g)$ solely from four-gluon amplitude.

For $\gamma_i$-deformed sYM theory the one-loop four-gluon amplitude is finite, so the $\beta$-function of the gauge coupling vanishes. 
This is consistent with the general formula
\begin{equation}
\beta(g)=\frac{Ng^3}{2}\left(-\frac{11}{3}+\frac{2n_f}{3}+\frac{n_s}{6}\right),
\end{equation}
where $n_f$ is the number of adjoint chiral fermions, and $n_s$ is the number of adjoint real scalars.

\section{The Flow of Yukawa Coupling\label{section:yukawa}}

In $\beta$-deformed sYM theory, the phase functions of Yukawa and scalar couplings are protected by the non-renormalization 
of superpotential. Without the protection of supersymmetry, the phase functions of $\gamma_i$-deformed sYM theory can receive 
both finite and infinite renormalization at the non-planar level. 
We will infer the consequences of the infinite renormalization using the strategy discussed in the previous section.

The $\beta$-deformed theory enjoys a $\mathbb{Z}_3$ symmetry which permutes the three chiral superfields and implies for the component 
action that all Yukawa couplings are equal. The different deformation parameters $\gamma_i$ of the $\gamma_i$-deformed sYM theory break 
this symmetry; thus, in principle, one can contemplate different values and renormalization constants for the Yukawa couplings. We will 
therefore further deform the scalar-fermion couplings of \eqref{gammaplanarl} to
\begin{equation}\label{lyukawa}
L_{Yukawa}^{\gamma}=\frac{-i}{\sqrt2}\sum_{A,B=1}^4h_{AB}e^{i\Theta_{AB}}
\Tr\Bigl[\bar{\phi}_{AB}\psi^A\psi^B+\phi^{AB}\bar{\psi}_A\bar{\psi}_B\Bigr],
\end{equation}
where $\Theta_{AB}\equiv \Theta(\psi^A\psi^B)$. The reality of $L_{Yukawa}^{\gamma}$ requires that $h$ is symmetric and $\Theta$ is 
antisymmetric in their indices, 
\begin{equation}
h_{AB}=h_{BA},\hspace{10mm}\Theta_{AB}=-\Theta_{BA} \ .
\end{equation}

Since $\gamma_i$-deformed sYM theory is conformal in the planar limit, quantum corrections corrections to $h_{AB}$ and $\Theta_{AB}$ 
can appear only at ${\cal O}(1/N^2)$:
\begin{equation}
\begin{aligned}
&h_{AB}=g+g\frac{k_{AB}}{N^2}+{\cal O}(g/N^4)~~~,\\
&\Theta_{AB}=\theta_{AB}+\frac{q_{AB}}{N^2}+{\cal O}(1/N^4),\\
\end{aligned}\label{htheta}
\end{equation}
where  $q_{AB}$ and $k_{AB}$ are functions of the $\gamma$ parameters. 

Since gauge coupling $g$ is not renormalized at one-loop level the flow of Yukawa couplings is solely induced by the flow 
of $\gamma_i$ parameters\footnote{If we were working in conventional renormalization procedures, $\gamma_i$'s should 
be dressed with $Z$ factors; these factors should become trivial in the supersymmetric limit $\gamma_1=\gamma_2=\gamma_3=\beta$.}. 
To use the method described in the previous section, to derive $\beta$-functions for the $\gamma_i$ parameters 
we need a sufficiently large set of on-shell tree amplitudes containing Yukawa couplings (and, for simplicity, preferably not involving 
scalar couplings), and the divergences of the corresponding one-loop amplitudes\footnote{The natural choice is three-point amplitude 
$\mathcal{A}(1\bar{\phi}_{AB}2\psi^A3\psi^B)$ , but for massless particles on shell condition requires momentum of three external 
legs to be parallel. One may use complex momenta in order to have non-zero three-point amplitudes.}. We choose the four-fermion 
amplitudes $\mathcal{A}(1\psi^A2\bar{\psi}_A3\psi^B4\bar{\psi}_B)$.

We extract the one-loop divergence of the amplitude $\mathcal{A}(1\psi^A2\bar{\psi}_A3\psi^B4\bar{\psi}_B)$ from the complete amplitude 
constructed though the generalized unitarity method \cite{Bern:1994zx, Bern:1996je, Britto:2004nc, Britto:2005ha}; 
to the order ${\cal O}(\frac{g^4}{N^2})$ it is
\begin{equation}
\begin{aligned}
&Div\Bigl(\mathcal{A}(1\psi^A2\bar{\psi}_A3\psi^B4\bar{\psi}_B)\Bigr)\\
=&\frac{g^4}{\epsilon N}A(1324)
\left(iD_I^{AB}\Bigl[\Tr_{\theta}(1324)-\Tr_{\theta}(1423)\Bigr]
+D_R^{AB}\Tr_{\theta}([1,3][2,4])\right) \ ,\\
\end{aligned}\label{fermiondiv}
\end{equation}
where $A(1324)$ is the color striped four-fermion tree amplitude without coupling constant. In the $\beta$-deformed sYM theory 
$D_I^{AB}=0$. The structure of the single-trace divergence in eq.~\eqref{fermiondiv} implies that cannot be removed by renormalizing 
the gauge of Yukawa couplings and thus a nonvanishing $D_I^{AB}$ indicates that the parameters $\gamma_i$ must be renormalized.

For $A=1, B=2$ the residues of the $1/\epsilon$ poles are given by
\begin{equation}
\begin{aligned}
D_R^{12}=&
-12k_{12}+8\sin\frac{\gamma_1+\gamma_2}{2}\Bigl(2+\cos(\gamma_1-\gamma_2)\Bigr),\\
D_I^{12}=&-2\sum_{B=1}^4q_{1B}+2\sum_{B=1}^4q_{2B}+2\Bigl(\sin2\gamma_1+\sin2\gamma_2
-2(\sin\gamma_1+\sin\gamma_2)\cos\gamma_3\Bigr),\\
\end{aligned}\label{div12}
\end{equation}
while for $A=3, B=4$ they are
\begin{equation}
\begin{aligned}
D_R^{34}=&-12k_{34}+8\sin\frac{\gamma_1-\gamma_2}{2}\Bigl(2+\cos(\gamma_1+\gamma_2)\Bigr),\\
D_I^{34}=&-2\sum_{B=1}^4q_{3B}+2\sum_{B=1}^4q_{4B}
+2\Bigl(-\sin2\gamma_1+\sin2\gamma_2+2(\sin\gamma_1-\sin\gamma_2)\cos\gamma_3\Bigr).\\
\end{aligned}\label{div34}
\end{equation}
The divergence of $\mathcal{A}(1\psi^A2\bar{\psi}_A3\psi^B4\bar{\psi}_B)$ for other choices of R-symmetry indices $A$ and  $B$ 
can be obtained by cyclically permuting the labels of fields and deformation parameters \footnote{The transformation \eqref{cyclic permutations} 
is equivalent to a cyclic permutation of $Q_i$'s in \eqref{qs}. There is another type of transformations of $Q_i$'s like $Q_1\rightarrow -Q_2,\ Q_2\rightarrow -Q_1,\ Q_3\rightarrow Q_3$, which relates $D^{ij}$ to $D^{i4}$. }:
\begin{equation}
\begin{aligned}
&\gamma_1\rightarrow \gamma_2,\ \gamma_2\rightarrow \gamma_3,\ \gamma_3\rightarrow \gamma_1,\\
&\psi^1\rightarrow \psi^2, \psi^2\rightarrow \psi^3, \psi^3\rightarrow \psi^1,\\
&\phi^1\rightarrow \phi^2, \phi^2\rightarrow \phi^3, \phi^3\rightarrow \phi^1.\\
\end{aligned}\label{cyclic permutations}
\end{equation}

To find the beta functions of the deformation parameters $\gamma_i$ we need the tree-level amplitudes 
$\mathcal{A}(1\psi^A2\bar{\psi}_A3\psi^B4\bar{\psi}_B)$. Up to ${\cal O}(\frac{g^4}{N^2})$ terms they are\footnote{In \eqref{fermiontree} double trace terms have been ignored since they are of order ${\cal O}(\frac{g^4}{N^2})$.}
\begin{equation}
\begin{aligned}
&\mathcal{A}^{tree}(1\psi^A2\bar{\psi}_A3\psi^B4\bar{\psi}_B)\\
=&A(1324)h_{AB}^2\Bigl(e^{2i\Theta_{AB}}\Tr(1324)+e^{-2i\Theta_{AB}}\Tr(1423)-\Tr(1342)
-\Tr(1243)\Bigr)\\
&+A(1234)g^2\Bigl(\Tr(1234)-\Tr(1243)-\Tr(1342)+\Tr(1432)\Bigr),\\
\end{aligned}\label{fermiontree}
\end{equation}
Since $\gamma_i$-deformed SYM is conformal at planar level, $\frac{d\gamma_m}{d\ln\mu}$ vanishes at the order ${\cal O}(\lambda)$,
\begin{equation}\label{dbetamorder}
\frac{d\gamma_m}{d\ln\mu}={\cal O}(\frac{\lambda}{N^2})={\cal O}(\frac{g^2}{N}) \ .
\end{equation}
Together with equations \eqref{htheta} and \eqref{dbetamorder} this implies that
\begin{equation}
\begin{aligned}
&\frac{dh_{AB}}{d\ln\mu}=\frac{g}{N^2}\frac{\partial k_{AB}}{\partial\gamma_m}\frac{d\gamma_m}{d\ln\mu}={\cal O}(\frac{g^3}{N^3}),\\
&\frac{d\Theta_{AB}}{d\ln\mu}=\frac{d\theta_{AB}}{d\ln\mu}+\frac{1}{N^2}\frac{\partial q_{AB}}{\partial\gamma_m}\frac{d\gamma_m}{d\ln\mu}=\frac{d\theta_{AB}}{d\ln\mu}+{\cal O}(\frac{g^2}{N^3}) \ .\\
\end{aligned}\label{dhdtheta}
\end{equation}
Thus, up to ${\cal O}(\frac{g^4}{N^2})$ terms, the derivative of the four-fermion tree-level amplitude with respect to $\ln\mu$ is
\begin{align}
\begin{split}
&\frac{d}{d\ln\mu}\mathcal{A}^{tree}(1\psi^A2\bar{\psi}_A3\psi^B4\bar{\psi}_B)
=2ig^2\frac{d\theta_{AB}}{d\ln\mu}A(1324)\Bigl(\Tr_{\theta}(1324)-\Tr_{\theta}(1423)\Bigr) \ .\\
\end{split}\label{dtree}
\end{align}

Finally, combining eqs.~\eqref{betadiv}, \eqref{fermiondiv} and \eqref{dtree} we find that the parameters of the theory
must be such that the following equations hold:
\begin{equation}
\begin{aligned}
0&=D_R^{AB},\\
\frac{d\theta_{AB}}{d\ln\mu}&=-\frac{g^2}{2N}D_I^{AB} \ .\\
\end{aligned}\label{renormeqs}
\end{equation}
In particular, these equations have no solution if $q_{AB}$ are set to 0, which suggests the finite renormalization $q_{AB}$ of $\theta_{AB}$ is nontrivial. They however do not uniquely fix $q_{AB}$ but rather it constrains them to be 
related to $\gamma_i$ as
\begin{equation}
\begin{aligned}
&q_{12}+q_{13}+q_{14}=\frac{1}{2}\sin(\gamma_1+\gamma_2)-\frac{1}{2}\sin(\gamma_1+\gamma_3)
+\frac{1}{2}\sin(-\gamma_2+\gamma_3),\\
&q_{21}+q_{23}+q_{24}=\frac{1}{2}\sin(\gamma_2+\gamma_3)-\frac{1}{2}\sin(\gamma_2+\gamma_1)
+\frac{1}{2}\sin(-\gamma_3+\gamma_1),\\
&q_{31}+q_{32}+q_{34}=\frac{1}{2}\sin(\gamma_3+\gamma_1)-\frac{1}{2}\sin(\gamma_3+\gamma_2)
+\frac{1}{2}\sin(-\gamma_1+\gamma_2) \ .\\
\end{aligned}\label{solq}
\end{equation}
To fix them uniquely it appears necessary to evaluate the ${\cal O}(\frac{g^4}{N^3})$ terms in the four-fermion amplitude. 
We will not do this explicitly here, but instead pick a natural solution to eq.~\eqref{solq},
\begin{equation}\label{naturalq}
q_{AB}=q(\theta_{AB})\equiv\frac{1}{2}\sin2\theta_{AB}-\frac{1}{2}\sin2\theta_{AB}|_{\gamma_m\rightarrow \bar{\gamma}} \ ,
\end{equation}
which has the correct ($q_{AB}=0$) $\mathcal{N}=1$ limit.

The same equations determine the finite renormalization $k_{AB}$ of $h_{AB}$:
\begin{equation}
\begin{aligned}
&k_{ij}=\frac{2}{3}\sin^2\frac{\gamma_i+\gamma_j}{2}\Bigl(2+\cos(\gamma_i-\gamma_j)\Bigr),\\
&k_{i4}=\frac{2}{3}\sin^2\frac{-\gamma_j+\gamma_k}{2}\Bigl(2+\cos(\gamma_j+\gamma_k)\Bigr) \ .\\
\end{aligned}\label{solk}
\end{equation}
In the $\mathcal{N}=1$ limit this solution reduces to 
\begin{equation}
\begin{aligned}
&k_{ij}=2\sin^2\beta,\ \ \ \ k_{i4}=0 \ .\\
\end{aligned}
\end{equation}
This limit is consistent with the relation \eqref{lbeta} 
\begin{equation}
\begin{aligned}
&h_{ij}=\frac{g}{\sqrt{1-\frac{4\sin^2\beta}{N^2}}},\ \ \ \ h_{i4}=g \\
\end{aligned}
\end{equation}
between $h_{AB}$, the gauge coupling and $\beta$  at the fixed point of the $\beta$-deformed sYM theory.

Last but not least, the second eq.~\eqref{renormeqs} also determined the $\beta$-functions of the $\gamma_i$ 
parameters. They are given by 
\begin{equation}
\begin{aligned}
&\frac{d\gamma_1}{d\ln\mu}=\frac{2g^2}{N}\sin\gamma_1(\cos\gamma_2+\cos\gamma_3-2\cos\gamma_1)\ ,\\
&\frac{d\gamma_2}{d\ln\mu}=\frac{2g^2}{N}\sin\gamma_2(\cos\gamma_3+\cos\gamma_1-2\cos\gamma_2)\ ,\\
&\frac{d\gamma_3}{d\ln\mu}=\frac{2g^2}{N}\sin\gamma_3(\cos\gamma_1+\cos\gamma_2-2\cos\gamma_3)\ .\\
\end{aligned}\label{gammaflow}
\end{equation}
We note that the eqs.~\eqref{gammaflow} are insensitive to the sign of $\gamma_i$ and thus, for the purpose of studying 
them\footnote{Since they are derived in a one-loop approximation these equations should be solved perturbatively in $g^2/N$. } 
we can assume that $0\le\gamma_m\le\pi$. Moreover, the right-hand side of these equations reflects the broken $\mathbb{Z}_3$
symmetry of the theory which may be restored by permuting $\gamma_i$.

The supersymmetric value of the deformation parameters,  $\gamma_1=\gamma_2=\gamma_3$, is a solution to these equations. Indeed,  
the right-hand side of \eqref{gammaflow} becomes trivial in this limit, reflecting the absence of renormalization of the deformation parameter.
In fact, the supersymmetric point acts as an attractor for flows starting sufficiently  close to it.  Indeed, linearizing around it reduces eqs.~\eqref{gammaflow} to
\begin{equation}
\begin{aligned}
&\frac{d\,\gamma_1}{d\ln\mu}=\frac{v}{3}(\;2\gamma_1-\gamma_2-\gamma_3)\ ,\\
&\frac{d\,\gamma_2}{d\ln\mu}=\frac{v}{3}(-\gamma_1+2\gamma_2-\gamma_3)\ ,\\
&\frac{d\,\gamma_3}{d\ln\mu}=\frac{v}{3}(-\gamma_1-\gamma_2+2\gamma_3) \ ;\\
\end{aligned}
\end{equation}
the coefficient 
\begin{equation}
v=\frac{6g^2\sin^2\bar{\gamma}}{N},
\label{def_v}
\end{equation}
characterizes the rate of the flow.

This system can be easily solved; defining 
\begin{equation}
\delta\gamma_i = \gamma_i -{\bar\gamma},\hspace{5mm}
\delta\gamma\equiv \sqrt{\delta\gamma_1^2+\delta\gamma_2^2+\delta\gamma_3^2}\ ,
\end{equation}
the solution is\footnote{Since $\delta\gamma_1+\delta\gamma_2+\delta\gamma_3=0$, \eqref{phasesolution} only has 2 independent equations.}
\begin{eqnarray}
\bar{\gamma}_i(\mu) &=&\bar{\gamma}_i(\mu_0)\ ,
\\
\delta\gamma_i(\mu) &=& \delta\gamma_i(\mu_0)\left(\frac{\mu}{\mu_0}\right)^{v}\ .
\label{phasesolution}
\end{eqnarray}
Since $v>0$, the second equation imply that far in the infrared the departures $\delta\gamma_i$ of all $\gamma_i$ parameter
from their average value become equal; thus if the flow of the deformation parameters starts close to a supersymmetric  point then they 
flow to a supersymmetric fixed point.

\begin{figure}
        \centering
        \begin{subfigure}[b]{0.45\textwidth}
                \includegraphics[scale=0.6]{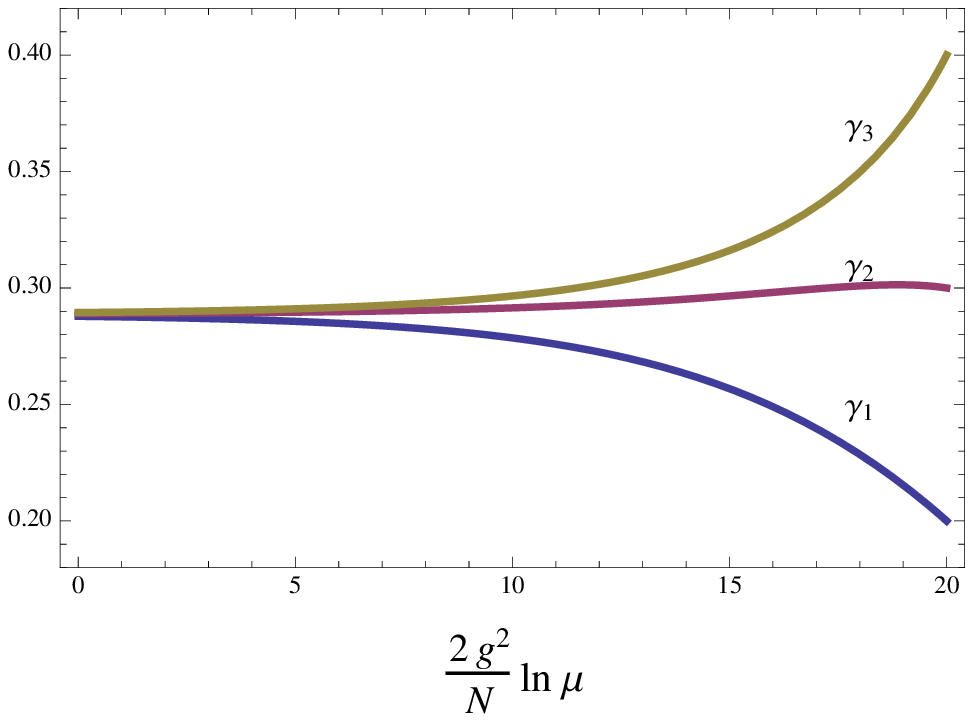}
                \caption{$\gamma_m^0=\{0.2,0.3,0.4\}$}
                \label{fig:betaflow1}
        \end{subfigure}
        \begin{subfigure}[b]{0.45\textwidth}
                \includegraphics[scale=0.6]{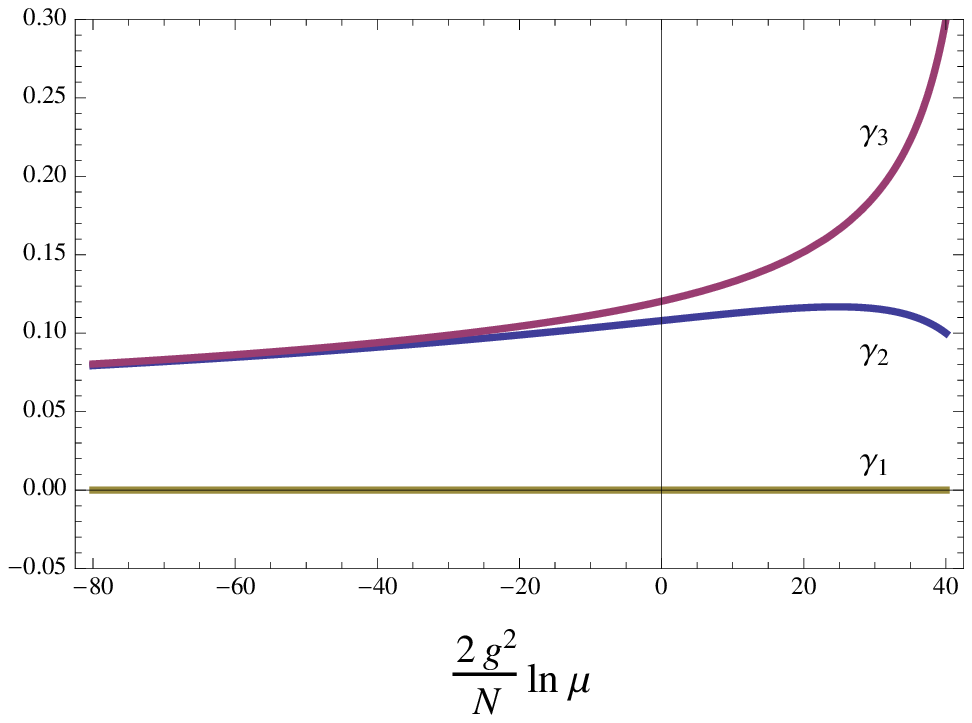}
                \caption{$\gamma_m^0=\{0.1,0.3,0\}$}
                \label{fig:betaflow2}
        \end{subfigure}
        \caption{The flow of phase with different initial values}\label{fig:betaflow}
\end{figure}

Far away from the $\mathcal{N}=1$ solution the flow of the deformation parameters can be analyzed by solving 
numerically system \eqref{gammaflow}. We show in Figure \ref{fig:betaflow1}, phases still flow to the same value; 
as in the linearized analysis all $\gamma_i$ parameters flow to a common value which is not necessarily the 
average $\bar{\gamma}$ of their initial values.

It is easy to see that the system \eqref{gammaflow} implies that if one of the parameters is vanishing at the 
renormalization scale then it remains zero along the flow. Solving the equations numerically reveals that 
the other parameters will flow to zero in the infrared; the resulting fixed point will exhibit $\mathcal{N}=4$ supersymmetry.
Figure \ref{fig:betaflow2} shows the case $\gamma_3=0$ and $\gamma_1\ne\gamma_2\ne 0$ at the renormalization scale; 
$\gamma_1$ and $\gamma_2$ will flow relatively rapidly to the same value and then together they slowly flow to zero.

A few limiting cases need a separate discussion.
The case one or more deformation parameters equal to $\pi$ is very similar to the case of one or more deformation parameters  
equal to 0. In the end all phases flow to $\gamma_i=\pi$ and the fixed-point theory is equivalent to $\mathcal{N}=4$ SYM. 
If all three $\gamma_i$ parameters are equal to 0 or $\pi$ then they do not change. 
If $0<\gamma_1<\pi$, $\gamma_2=0$, $\gamma_3=\pi$ only the first equation \eqref{gammaflow} is nontrivial and  
can be solved analytically:
\begin{equation}
\gamma_1(\mu)=\cot^{-1}\left(\cot\gamma_1(\mu_0)\left(\frac{\mu}{\mu_0}\right)^{4g^2/N}\right) \ ,
\end{equation}
implying that and $\gamma_1$ will flow to
\begin{equation}
\gamma_1(0)=\frac{\pi}{2} \ .
\end{equation}

Although, as we have seen, the Yukawa couplings flow to an ${\cal N}=1$ infrared fixed point almost in the entire parameter space, 
in the next section we will show that scalar couplings have a similar property only if $\delta\gamma\sim {\cal O}(\frac{1}{N^2})$, that is 
only if the flow begins very close to a supersymmetric point.  Thus,  ${\cal O}(\delta\gamma^2)$ terms can be discarded in the Lagrangian 
if we are only keeping ${\cal O}(\frac{1}{N^2})$ terms,
\begin{equation}
\begin{aligned}
h_{ij}&=g\left(1+\frac{2}{N^2}\sin^2\frac{\gamma_i+\gamma_j}{2}\right)\sim h_0(\theta_{ij}),\\
h_{i4}&=g,\\
\end{aligned}\label{yukawaorder1}
\end{equation}
where $h_0$ as a function of $g$ and phase was defined in eq.~\eqref{h0} and $\theta_{ij}\equiv\theta(\phi^i\phi^j)$.
The Lagrangian of Yukawa terms including terms of order ${\cal O}(\frac{1}{N^2})$ is
\begin{equation}
\begin{aligned}
L_{Yukawa}^\gamma=\Tr&\Bigl[-\frac{ih_0(\theta_{ki})}{\sqrt{2}}\left(\epsilon^{ijk}\bar{\psi}_i\left[\bar{\phi}_j,\ \bar{\psi}_k\right]_{\Theta}+\epsilon_{ijk}\psi^i[\phi^j,\ \psi^k]_{\Theta}\right)\\
&+\sqrt{2}ig\left(\bar{\psi}_i\left[\phi^{i},\ \bar{\psi}_4\right]_{\Theta}+\psi^i\left[\bar{\phi}_{i},\ \psi^4\right]_{\Theta}\right)\Bigr]\\
\end{aligned}\label{eq:yukawal one over n square}
\end{equation}

\section{The Flow of Scalar Coupling\label{section:scalar}}

As discussed in detail in \cite{Fokken:2013aea}, renormalizability of the four-scalar terms in the $\gamma_i$-deformed theory requires that  
the action be extended with double-trace terms at order ${\cal O}(\lambda/N)={\cal O}(g^2)$  and further single trace terms at order 
${\cal O}(g^2/N)$. Adding all such terms the four-scalar Lagrangian becomes
\footnote{In principle one can also add $\Tr(\phi^i\bar{\phi}_i)\Tr(\phi^j\bar{\phi}_j)$ terms to \eqref{phi4} but, 
at least at one-loop level, they are not required by renormalizability.},
\begin{equation}
\begin{aligned}
L_{4\phi}^{\gamma}&=g^2\Tr\left([\phi^i,\ \phi^j]_{\Theta}[\bar{\phi}_i,\ \bar{\phi}_j]_{\Theta}-
\frac{1}{2}[\phi^i,\ \bar{\phi}_i]^2\right)\\
+&\frac{2}{N^2}\left(a_1^{ij}\Tr_{\theta}(\phi^i\phi^j\bar{\phi}_i\bar{\phi}_j)
+a_2^{ij}\Tr_{\theta}(\phi^i\phi^j\bar{\phi}_j\bar{\phi}_i)
+a_3^{ij}\Tr_{\theta}(\phi^i\bar{\phi}_i\phi^j\bar{\phi}_j)\right)\\
+&\frac{1}{N^2}\left(a_4^i\Tr(\phi^i\phi^i\bar{\phi}_i\bar{\phi}_i)
+a_5^i\Tr(\phi^i\bar{\phi}_i\phi^i\bar{\phi}_i)\right)
+\frac{1}{N}a_{6}^i\Tr(\phi^i\phi^i)\Tr(\bar{\phi}_i\bar{\phi}_i)\\
+&\frac{1}{N}\left(a_7^{ij}\Tr(\phi^i\phi^j)\Tr(\bar{\phi}_i\bar{\phi}_j)
+a_8^{ij}\Tr(\phi^i\bar{\phi}_j)\Tr(\bar{\phi}_i\phi^j)\right) \ ; \\
\end{aligned}\label{phi4}
\end{equation}
on the first line the phase functions are defined by
\begin{equation}
\begin{aligned}
&\Theta_{ij}^{\phi}\equiv\Theta(\phi^i\phi^j),\\
&\Theta_{ij}^{\phi}=\theta_{ij}^{\phi}+\frac{q^{\phi}_{ij}}{N^2},\ \ \ \theta_{ij}^{\phi}=\epsilon_{ijk}\gamma_k.\\
\end{aligned}
\end{equation}

To derive beta functions of these $a_I$ factors, we need to compute four-scalar amplitude at tree and one-loop level. 
Explicit calculations show that the one-loop divergences of such amplitudes have the following structure:
\begin{eqnarray}
&&Div(\mathcal{A}(1\phi^i2\bar{\phi}_i3\phi^j4\bar{\phi}_j)=\frac{iI_{ij}}{\epsilon N}\Bigl(\Tr_{\theta}(1324)-\Tr_{\theta}(1423)\Bigr)
\nonumber\\
&&+\frac{1}{\epsilon N}\left(D_2^{ij}(\Tr(1342)+\Tr(1243)
+D_3^{ij}(\Tr(1234)+\Tr(1432))\right)                       \label{1stphi4}\\
&&+\frac{1}{\epsilon N}D_1^{ij}\Bigl(\Tr_{\theta}(1324)+\Tr_{\theta}(1423)\Bigr)
+D_7^{ij}\Tr(13)\Tr(24)+D_8^{ij}\Tr(14)\Tr(23) \ ,
\nonumber
\\[5pt]
&&Div(\mathcal{A}(1\phi^i2\bar{\phi}_i3\phi^i4\bar{\phi}_i)
=\frac{1}{\epsilon N^2}D_4\Bigl(\Tr(1324)+\Tr(1342)+\Tr(1423)+\Tr(1243)\Bigr)
\nonumber\\
&&+\frac{1}{\epsilon N^2}D_5^i\Bigl(\Tr(1234)+\Tr(1432)\Bigr)+\frac{1}{\epsilon N}D_{6}^i\Tr(13)\Tr(24) \ .
\label{2ndphi4}
\end{eqnarray}
The discussion in sec.~\ref{section:betafromdivergence} and in particular eq.~\eqref{betadiv}, imply that the $\beta$-functions 
of the parameters $a_I$  and of the deformation $q_{ij}$ of the phase function are given by
\begin{equation}
\begin{aligned}
\frac{da_I}{d\ln\mu}&=-ND_I,\\
\frac{dq^{\phi}_{ij}}{d\ln\mu}&=-\frac{N}{2}I_{ij}.\\
\end{aligned}\label{di}
\end{equation}
Before find the explicit expressions for $D_I$ and $I_{ij}$ we note that, since $a_1-a_5$ always appear accompanied 
by a $\frac{1}{N^2}$ factor,  the divergence requiring the presence of the double-trace terms cannot depend on single-trace terms. 
Thus, the renormalization group (RG) flow equations of double trace operators is a subsystem of \eqref{di}. We will solve this subsystem in subsection 
\ref{double}, and solve the remaining equations in subsection \ref{single}.

\subsection{Flow of Double-Trace operators \label{double}}

Apart form the gauge coupling, the divergent double-trace terms in the four-scalar amplitudes depend only on $a_6$, $a_7$, $a_8$
and the deformation parameters $\gamma_i$; the relevant coefficients in eqs.~\eqref{1stphi4} and \eqref{2ndphi4} are given by
\begin{equation}
\begin{aligned}
D_{6}^1&=\frac{1}{2}(a_{6}^1)^2+32g^4(\sin^2\gamma_2-\sin^2\gamma_3)^2 \ ,\\
D_7^{23}&=(a_7^{23})^2-8g^2\sin^2\gamma_1a_7^{23}
+16g^4\left(\sin^2\gamma_1-\sin^2\frac{\gamma_2-\gamma_3}{2}\right)^2 \ ,\\
D_8^{23}&=(a_8^{23})^2-8g^2\sin^2\gamma_1a_8^{23}
+16g^4\left(\sin^2\gamma_1-\sin^2\frac{\gamma_2+\gamma_3}{2}\right)^2 \ .\\
\end{aligned}\label{doubletrace}
\end{equation}
We note that, since each equation depends on a single double-trace deformation parameter, their flows are independent of each other.

The solution for $a_{6}^{1}$ is
\begin{equation}
\begin{aligned}
a_{6}^1&=-g^2A_{23}\tan\left(\frac{g^2N A_{23}}{2}\ln\frac{\mu}{\mu_1}\right),\\
A_{23}&=8|\sin^2\gamma_2-\sin^2\gamma_3|,\\
T_6^1&=\frac{g^2N A_{23}}{2}\ln\frac{\mu}{\mu_1},\\
\end{aligned}\label{a6solution}
\end{equation}
where $\mu_1$ is the energy scale at which $a_6^1=0$ and $T_6^1$ is introduced for future notational convenience. 
If $\mu>\mu_1$, $a_6^1(\mu)<0$, which renders the potential unbounded from below and thus destabilizes the theory; to avoid such situations
we need to require that $\mu_0\le\mu_1$. Since $D_6^1$ is always positive, if the $\gamma_i$ parameters were fixed then $a_6^1$ would 
inevitably flow to positive infinity in the infrared. 
Fortunately however, as discussed in the previous section, the $\gamma_i$ parameters vary with the scale as $\mu^v$ with $v$ defined in eq.~\eqref{def_v}. Their running suppresses the logarithmic running of $a_6^1$.When $\mu\rightarrow 0$ and implies that
\begin{equation}
a_{6}^1= {\cal O}(g^4N\delta\gamma^2\ln\mu) \ .
\end{equation}

\begin{figure}
        \centering
        \begin{subfigure}[b]{0.45\textwidth}
                \includegraphics[scale=0.6]{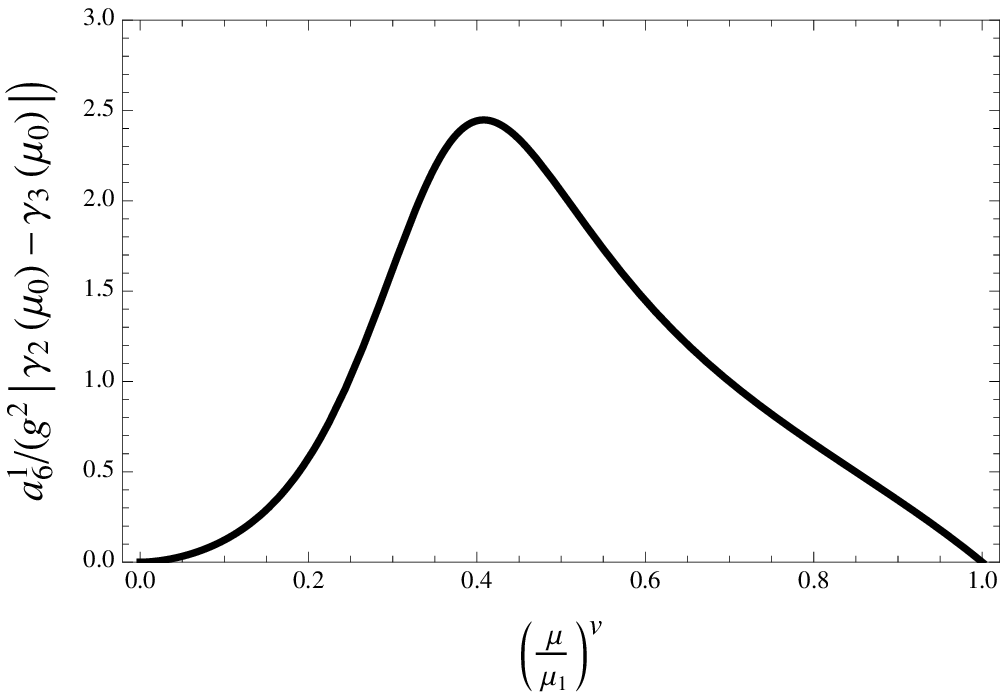}
                \caption{$|\gamma_2(\mu_0)-\gamma_3(\mu_0)|=0.9\frac{3e\pi|\tan\bar{\gamma}|}{8N^2}$}
                \label{fig:a6flow1}
        \end{subfigure}
        \begin{subfigure}[b]{0.45\textwidth}
                \includegraphics[scale=0.6]{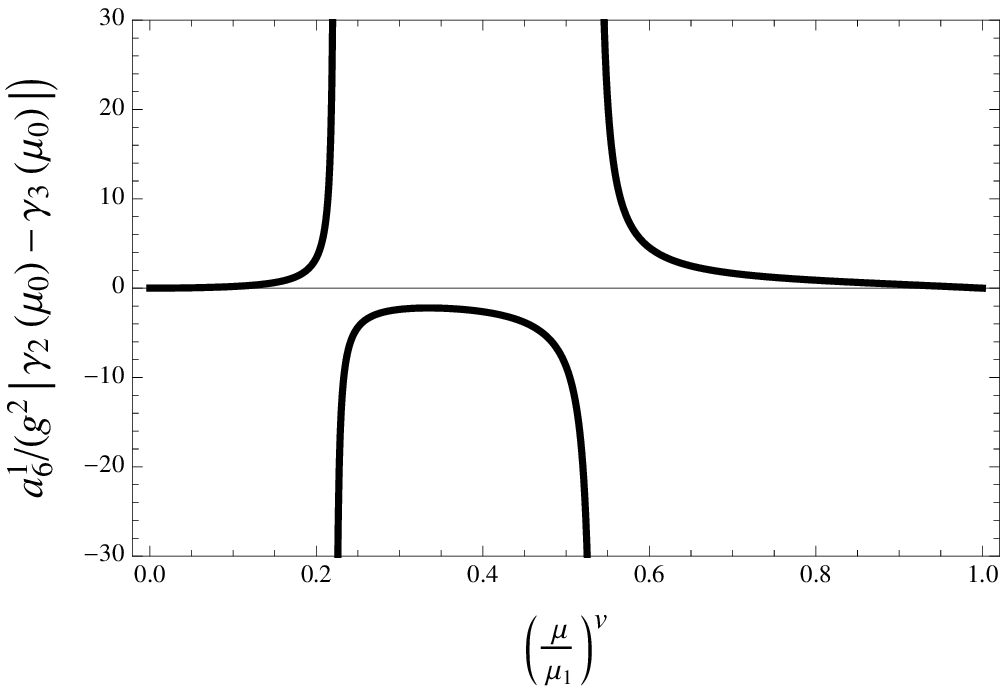}
                \caption{$|\gamma_2(\mu_0)-\gamma_3(\mu_0)|=1.1\frac{3e\pi|\tan\bar{\gamma}|}{8N^2}$}
                \label{fig:a6flow2}
        \end{subfigure}
        \caption{The flow of $a_6^{1}$ with different $|\gamma_2^0-\gamma_3^0|$}\label{fig:a6flow}
\end{figure}

A potential issue with this argument is that $a_6^1$ may run to infinity before it reaches the infrared fixed point. 
For $a_6^1$ to be well defined at any energy scale $\mu<\mu_0$, the argument   $T_6^1$ of the tangent function in 
eq.~eqref{a6solution} must satisfy $|T_6^1|<\frac{\pi}{2}$ for any  $\mu<\mu_0$. 
To explore this issue let us define $\gamma_m(\mu_1)\equiv \gamma_m^1$ and expand the solution to the RG flow 
equations around the supersymmetric point;
\begin{equation}
\begin{aligned}
A_{23}&=8|\sin2\bar{\gamma}(\gamma_2-\gamma_3)|,\\
T_6^1&=4g^2|\sin2\bar{\gamma}\left(\gamma_2^1-\gamma_3^1\right)|(\frac{\mu}{\mu_1})^v\ln\frac{\mu}{\mu_1}.\\
\end{aligned}\label{T61}
\end{equation}
where we have used eq.~\eqref{phasesolution} for the behavior of $\delta\gamma_i$ around the fixed point. 
The maximum of $T_6^1$ is at $\mu_M=\mu_1e^{-\frac{1}{v}}$.  If $\mu_0\ge\mu_M$, $a_6^1$ will run to infinity unless
\begin{equation}\label{con1}
\left|\frac{g^2N A_{23}(\mu_M)}{2}\ln\frac{\mu_M}{\mu_1}\right|<\frac{\pi}{2} \ ;
\end{equation}
This inequality constrains $\delta\gamma_m^1$ and $\delta\gamma_m(\mu_0)$ to be quite close to a supersymmetric point:
\begin{equation}
|\gamma_2(\mu_0)-\gamma_3(\mu_0)|\le |\gamma_2^1-\gamma_3^1|<\frac{3e\pi|\tan\bar{\gamma}|}{8N^2} \ .
\end{equation}
In Figure \ref{fig:a6flow1} and Figure \ref{fig:a6flow2} we plot the graph of $a_6^1$ as a function of $\mu$ for 
$|\gamma_2(\mu_0)-\gamma_3(\mu_0)|$ less and greater than $\frac{3e\pi|\tan\bar{\gamma}|}{8N^2}$, respectively. 

A logical possibility is that although $a_6^1$ may not be well defined in the entire energy $\mu<\mu_M$ range, it is well 
defined in a region $\mu\le \mu_0$ with $\mu_0<\mu_M$, corresponding to the top-left corner of Figure \ref{fig:a6flow2}. 
In this case the $T$ parameter introduced in eq.~\eqref{doubletrace} is given by 
\begin{equation}\label{con2}
|T_6(\mu_0)|=\left|\frac{g^2N A_{23}(\mu_0)}{2}\ln\frac{\mu_0}{\mu_1}\right|<\frac{\pi}{2} \ ;
\end{equation}
This inequality constrains $\delta\gamma_m^0$ to obey the equation,
\begin{equation}
|\gamma_2(\mu_0)-\gamma_3(\mu_0)|<\left|\frac{\pi}{8g^2N \sin2\bar{\gamma}\ln\frac{\mu_0}{\mu_1}}\right|
<\left|\frac{\pi v}{8g^2N \sin2\bar{\gamma}}\right|=\frac{3\pi|\tan\bar{\gamma}|}{8N^2 }.
\end{equation}
Thus, in both cases, $a_6$ runs to infinity over some finite energy interval unless unless the $\gamma_i={\bar\gamma}+\delta\gamma_i$
parameters are very close to their supersymmetric values,  
\begin{equation}\label{deltabeta}
\delta\gamma\sim {\cal O}(1/N^2) \ .
\end{equation}

Next let up consider the running of the other double-trace terms. The solution for $a_7^{23}$ to the RG flow equations \eqref{doubletrace} is
\begin{equation}
\begin{aligned}
&a_7^{23}=g^2A_7\pm g^2\sqrt{A_7^2-B_7^2}\tanh\left(g^2N\sqrt{A_7^2-B_7^2}\ln\frac{\mu}{\mu_7}\right),\\
&A_7=4\sin^2\gamma_1,\hspace{3mm}B_7=4\left(\sin^2\gamma_1-\sin^2\frac{\gamma_2-\gamma_3}{2}\right),\\
\end{aligned}
\end{equation}
where $\mu_7$ is a constant which can be fixed by choosing boundary conditions ({\it i.e.} by choosing a renormalization condition). 
When $\mu\rightarrow 0$, $A_7^2-B_7^2 \sim \mu^{2v}$,
\begin{equation}
a_7^{23}=4g^2\sin^2\gamma_1+ {\cal O}(g^4N\delta\gamma^2\ln\mu) \ .
\end{equation}
As a comparison, in the $\beta$-deformed Lagrangian \eqref{lbeta}, the counterpart of $a_7$ is contained in the last line,
\begin{equation}
\begin{aligned}
&h^2\Tr\left([\phi^{i},\ \phi^{j}]_{\theta}T^a\right)\Tr\left([\bar{\phi}_{i},\ \bar{\phi}_{j}]_{\theta}T^a\right)\\
\rightarrow& -\frac{h^2}{N}\Tr\left([\phi^{i},\ \phi^{j}]_{\theta}\right)\Tr\left([\bar{\phi}_{i},\ \bar{\phi}_{j}]_{\theta}\right)=\frac{4g^2}{N}\sin^2\beta\Tr\left(\phi^{i} \phi^{j}\right)\Tr\left(\bar{\phi}_{i}\bar{\phi}_{j}\right)
\end{aligned}
\end{equation}

Last but not least, the RG flow of the coefficient $a_8^{23}$ is
\begin{equation}
\begin{aligned}
a_8^{23}&=g^2A_8+g^2\sqrt{A_8^2-B_8^2}\tanh\left(g^2N\sqrt{A_8^2-B_8^2}\ln\frac{\mu}{\mu_8}\right),\\
A_8&=4\sin^2\gamma_1,\hspace{3mm}B_8=4\left(\sin^2\gamma_1-\sin^2\frac{\gamma_2+\gamma_3}{2}\right) \ ,\\
\end{aligned}
\end{equation}
where, similarly to the previous case, $\mu_8$ is a constant which can be fixed by boundary conditions. 
When $\mu\rightarrow 0$, $B_8^2 \rightarrow 0$, $\tanh\left(g^2N\sqrt{A_8^2-B_8^2}\ln\frac{\mu}{c}\right)\rightarrow -1$,
\begin{equation}
a_8^{23}= {\cal O}(g^2\delta\gamma^2) \ .
\end{equation}

To summarize this calculation, up to terms of the order of ${\cal O}(\delta\gamma^2)$, the RG evolution of the double-trace  deformations
of the Lagrangian is
\begin{equation}\label{doubletracesolution}
a_6^i=a_8^{ij}=0,\hspace{5mm}a_7^{ij}=4\sin^2\theta_{ij}^{\phi} \ .
\end{equation}
The restriction to only terms of order ${\cal O}(\delta\gamma)$ is justified by the requirement starts close to (but {\it not} at) a 
supersymmetric configuration such that running double-trace couplings are well-defined for all energy scales.

\subsection{Flow of Single-Trace Operators\label{single}}

Let us now discuss the flow of the couplings $a_1,\  a_2\cdots a_5$ characterizing the single-trace four-scalar terms.
The divergences $D_1,\  D_2\cdots D_5$ of the four-scalar amplitudes are given explicitly in Appendix~\ref{appendixA}, 
and they can be written compactly in the form
\begin{equation}
D_I=-g^2(M_{IJ}a_J+V_I) \ .
\end{equation}
Here the index  $I$ runs over $I=1,\cdots, 5$, $M_{IJ}$ is a constant matrix whose eigenvalues\footnote{The eigenvalues of $M_{IJ}$ are 
$\{\frac{13+\sqrt{41}}{2},\ \frac{13-\sqrt{41}}{2},\ 5+\sqrt{5},\ 5+\sqrt{5},\ 5-\sqrt{5},\ 5-\sqrt{5},\ 6,\ 6,\ 6,\ 6,\ 6,\ 6,\ 0,\ 0,\ 0\}$.} are all non-negative 
and the vector  $V_I$ contains double-trace couplings $a_{6, 7, 8}$ which may be replaced with their solutions in \eqref{doubletracesolution}.

The matrix $M_{IJ}$ has a three-dimensional null space; the corresponding orthonormal eigenvectors $\{n_1,n_2,n_3\}$ form 
the basis of this space; using them we can construct three combinations of couplings, $e_i=(n_i)_Ja_J$,  whose running is trivial. Indeed, 
$e_i$ are given by
\begin{equation}
\begin{aligned}
&e_1=a_1^{23}+a_2^{23}+a_3^{23}-\frac{1}{2}(a_4^2+a_4^3)-\frac{1}{4}(a_5^2+a_5^3),\\
&e_2=a_1^{31}+a_2^{31}+a_3^{31}-\frac{1}{2}(a_4^3+a_4^1)-\frac{1}{4}(a_5^3+a_5^1),\\
&e_3=a_1^{12}+a_2^{12}+a_3^{12}-\frac{1}{2}(a_4^1+a_4^2)-\frac{1}{4}(a_5^1+a_5^2).\\
\end{aligned}\label{ei}
\end{equation}
and it is not difficult to check that the inner-product $n_i\cdot V = \sum_I(n_i)_IV_I$ vanishes up to terms of the order of 
${\cal O}(\delta\gamma^2)$, 
\begin{equation}
\begin{aligned}
n_i^IV_I&={\cal O}(\delta\gamma^2),\\
\frac{d e_i}{d\ln\mu}&=n_i^IM_{IJ}a_J+n_i^IV_I=n_i^IV_I={\cal O}(\delta\gamma^2) \ .\\
\end{aligned}
\end{equation}
Thus, up to ${\cal O}(\delta\gamma^2)$ terms, we can set $e_i=0$ and eliminate $a_5$ using eqs.~\eqref{ei}. 
The new matrix $M_{IJ}$ describing the RG evolution of the remaining couplings  has only positive eigenvalues. 
The solution to the matrix RG flow equations
\begin{equation}
\frac{d a_I}{d \ln\mu}=Ng^2(M_{IJ}a_J+V_I),
\end{equation}
can be written as
\begin{equation}\label{singletracesolution}
a_I=Ng^2(\mu^{Ng^2M})_{IJ}\int d\ln\mu (\mu^{-Ng^2M}V)_J=-(M^{-1}V)_I+Ng^2(\mu^{Ng^2M})_{IJ}c_J,
\end{equation}
where $c_J$ can be fixed by boundary conditions. Since the eigenvalues of $M_{IJ}$ are ${\cal O}(1)$ the second term 
in \eqref{singletracesolution} goes to zero in the infrared ($\mu\rightarrow 0$) much faster than $\delta\gamma^2$, so at 
low energies
\begin{equation}
a_I\rightarrow -(M^{-1}V)_I \ .
\end{equation}
Using the explicit expression of the matrix $M_{IJ}$ and of the vector $V_I$ we find that the running couplings $a_{1,2,3,4,5}$ are 
related to the running $\delta\gamma_i$ as
\begin{equation}
\begin{aligned}
&a_1^{ij}=-a_2^{ij}=4g^2\sin^2\theta^{\phi}_{ij},\\
&a_3^{ij}=a_4^i=a_5^i=0 \ .\\
\end{aligned}
\label{singletrace}
\end{equation}

Last, we discuss the renormalization of the phase functions $\theta_{ij}$. Using the $\beta$-functions $\frac{d\gamma_m}{d\ln\mu}$ 
for the $\gamma_i$-parameters and $I_{ij}$ included in eq.~\eqref{i12} in Appendix~\ref{appendixA} in the second line of eq.~\eqref{di} 
we find that up to ${\cal O}(\delta\gamma^2)$ terms,
\begin{equation}
q_{ij}^{\phi}=\frac{1}{2}\sin2\theta_{ij}^{\phi}-\frac{1}{2}\sin2\theta_{ij}^{\phi}|_{\gamma_m\rightarrow \bar{\gamma}} \ .
\end{equation}
Thus, the renormalization of the phase function $\theta_{ij}$ is quite similar to $q_{AB}$ in eq.~\eqref{naturalq} which 
renormalizes the phase function in the Yukawa couplings.

\subsection{The Emergence of Supersymmetry}

We are now in position to assemble the infrared four-scalar Lagrangian while accounting for the flow of the couplings. 
Using the solutions for the $a_I$ parameters in eqs.~\eqref{doubletracesolution} and \eqref{singletrace} in the Lagrangian 
\eqref{phi4} we find that
\begin{equation}
\begin{aligned}
L_{4\phi}^{\mathcal{N}=0}&=g^2\Tr\left([\phi^i,\ \phi^j]_{\Theta}[\bar{\phi}_i,\ \bar{\phi}_j]_{\Theta}-
\frac{1}{2}[\phi^i,\ \bar{\phi}_i]^2\right)\\
+&\sum_{i\ne j}\frac{4g^2\sin^2\theta_{ij}^{\phi}}{N^2}\Bigl(2\Tr_{\theta}(\phi^i\phi^j\bar{\phi}_i\bar{\phi}_j)
-2\Tr_{\theta}(\phi^i\phi^j\bar{\phi}_j\bar{\phi}_i)+N\Tr(\phi^i\phi^j)\Tr(\bar{\phi}_i\bar{\phi}_j)\Bigr)\\
&\sim \frac{g^2}{1-\frac{4\sin^2\theta_{ij}^{\phi}}{N^2}}\Tr([\phi^i,\ \phi^j]_{\Theta}T^a)\Tr([\bar{\phi}_i,\ \bar{\phi}_j]_{\Theta})-\frac{g^2}{2}\Tr\left([\phi^i,\ \bar{\phi}_i]^2\right).\\
\end{aligned}\label{phi4order1}
\end{equation}

Adding the gauge and Yukawa interactions, the complete Lagrangian of $\gamma_i$-deformed sYM theory up to ${\cal O}(\frac{1}{N^2})$ 
terms  is 
\begin{equation}
\begin{aligned}
L_\gamma=\Tr&\Bigl[-\frac{1}{4}F_{\mu\nu}F^{\mu\nu}-D_{\mu}\bar{\phi}_{i} D^{\mu}\phi^{i}
+i\bar{\psi}_i\bar{\sigma}^{\mu}D_{\mu}\psi^i+i\bar{\psi}_4\bar{\sigma}^{\mu}D_{\mu}\psi^4\\
&-\frac{g^2}{2}[\phi^{i},\ \bar{\phi}_{i}]^2-\frac{ih_0(\theta_{ki})}{\sqrt{2}}\left(\epsilon^{ijk}\bar{\psi}_i\left[\bar{\phi}_j,\ \bar{\psi}_k\right]_{\Theta}+\epsilon_{ijk}\psi^i[\phi^j,\ \psi^k]_{\Theta}\right)\\
&+\sqrt{2}ig\left(\bar{\psi}_i\left[\phi^{i},\ \bar{\psi}_4\right]_{\Theta}+\psi^i\left[\bar{\phi}_{i},\ \psi^4\right]_{\Theta}\right)\Bigr]\\
+h_0^2&(\theta^{\phi}_{ij})\Tr\Bigl[[\phi^{i},\ \phi^{j}]_{\Theta}T^a\Bigr]\Tr\Bigl[[\bar{\phi}_{i},\ \bar{\phi}_{j}]_{\Theta}T^a\Bigr] \ .\\
\end{aligned}\label{eq:gammal}
\end{equation}
Here the phase functions are defined as
\begin{equation}
\Theta(\Phi_1\Phi_2)=\theta(\Phi_1\Phi_2)+\frac{1}{2N^2}\left(\sin\theta(\Phi_1\Phi_2)
-\sin\theta(\Phi_1\Phi_2)|_{\gamma_m\rightarrow \bar{\gamma}}\right) 
\end{equation}
that is, they include one-loop corrections to the tree-level phase functions.

As discussed in sec.~\ref{section:yukawa}, the deformation parameters $\gamma_i$ generically flow in the infrared to a 
supersymmetric configurations, in which they all become equal and the phase functions become
\begin{equation}
\Theta_{ij}\rightarrow \pm\beta ,\hspace{7mm}
\Theta_{i4}\rightarrow 0 ,\hspace{7mm}
\Theta_{ij}^{\phi}\rightarrow \pm\beta \ ;
\end{equation}
therefore, $L_\gamma$  reduces to the Lagrangian \eqref{lbeta} of the $\beta$-deformed sYM theory.

\section{Discussion\label{section:discussion}}

The $\gamma_i$-deformed $\mathcal{N}$=4 sYM theory is a non-supersymmetric deformation of maximally supersymmetric 
gauge theory in four dimensions which is conformally invariant in the planar limit. It has been argued \cite{AD_RR_up, Fokken:2013aea}
that, generically, one-loop-induced double-trace operators destabilize the theory by rendering its potential unbounded from below 
through a renormalization group flow.
In this paper we analyzed in details the one-loop renormalization of the complete theory and the ensuing renormalization group flow of 
all Yukawa and four-scalar couplings and found that if the flow starts close to a supersymmetric configuration of parameters -- 
$\gamma_i={\bar\gamma}+\delta\gamma_i$ with $\delta\gamma_i={\cal O}(1/N^2)$ -- then the theory remains stable throughout the flow 
and reaches a fixed point in the far infrared. The fixed-point theory is the ${\cal N}=1$-supersymmetric $\beta$-deformed $\mathcal{N}=4$ 
sYM theory.

It would be interesting to explore whether similar phenomena occur in other field theories;
an obvious candidate is the  $\gamma_i$-deformed $\mathcal{N}$=4 sYM theory with complex $\gamma_i$ which may be expected 
to flow to the $\beta$-deformed $\mathcal{N}$=4 sYM theory with a complex $\beta$ parameter.




The running of marginal scalar double-trace operators is also a feature of abelian orbifolds of $\mathcal{N}=4$ SYM. In the dual string theory context, the running corresponds to the presence of closed string tachyons at weak coupling \cite{Dymarsky:2005uh,Dymarsky:2005nc}, and a non-perturbative instability at strong coupling \cite{Horowitz:2007pr}. In $\gamma_i$-deformed SYM, the running is suppressed by the flow of phase parameters $\gamma_i$. We expect that, if this effect has a counterpart in the large volume limit of the dual string theory, it is presumably related to
an instanton similar to the one discussed in \cite{Horowitz:2007pr} for non-supersymmetric orbifolds of AdS$_5 \times $S$^5$.

\

\section*{Acknowledgements}

I would like to express my sincere gratitude to my advisor, Radu Roiban, for valuable discussions and comments and for revising the manuscript. This work is supported in part by the  US DoE under contract DE-SC0008745.

\appendix
\numberwithin{equation}{section}
\section{Single Trace Divergence of $\phi^4$ Amplitude\label{appendixA}}

In this appendix we collect the residues of the single-trace divergences of one-loop four-scalar amplitudes 
used in sec.~\ref{single}:
\begin{equation}
\begin{aligned}
I_{12}/g^2=&4g^2(q_{14}+q_{42}+q_{23}+q_{31}-2q_{12}^{\phi})
-\sin4\gamma_3(a_{6}^1+a_{6}^2)+\sin2(\gamma_2-\gamma_3)a_7^{23}\\
&+\sin2(\gamma_1-\gamma_3)a_7^{13}
-\sin2(\gamma_2+\gamma_3)a_8^{23}-\sin2(\gamma_1+\gamma_3)a_8^{13}\\
\end{aligned}\label{i12}
\end{equation}
\begin{align}
\begin{split}
D_1^{12}/g^2=&4g^2\Bigl(2
-\cos(\gamma_2+\gamma_3)-\cos(\gamma_2-\gamma_3)-\cos(\gamma_1+\gamma_3)-
\cos(\gamma_1-\gamma_3)\\
&\hspace{7mm}-2\cos\gamma_3+2\cos(\gamma_1+\gamma_2)+
2\cos(\gamma_1-\gamma_2)\Bigr)\\
&-4a_1^{12}+2a_2^{12}+2a_3^{12}+\cos4\gamma_3(a_{6}^1+a_{6}^2)+
\cos2(\gamma_1-\gamma_3)a_7^{13}\\
&+\cos2(\gamma_2-\gamma_3)a_7^{23}+\cos2(\gamma_1+\gamma_3)a_8^{13}
+\cos2(\gamma_2+\gamma_3)a_8^{23}\\
\end{split}\\
\begin{split}
D_2^{12}/g^2=&2g^2\Bigl(5+\cos2\gamma_1+\cos2\gamma_2+4\cos2\gamma_3
-2\cos2(\gamma_1-\gamma_2)\\
&\hspace{7mm}+4\cos(\gamma_1-\gamma_2-2\gamma_3)+4\cos(\gamma_1
-\gamma_2+2\gamma_3)\Bigr)\\
&-2g^2\left(k_{13}+k_{23}+k_{14}+k_{24}\right)+2a_1^{12}-\frac{1}{2}(a_2^{13}+a_2^{23})-5a_2^{12}\\
&-\frac{1}{2}(a_3^{13}+a_3^{23})
+a_3^{12}-\frac{1}{2}(a_{4}^1+a_{4}^2)-\frac{1}{2}(a_{5}^1+a_{5}^2)-\frac{1}{2}(a_{6}^1+a_{6}^2)\\
&-\frac{1}{2}(a_7^{13}+a_7^{23})-3a_7^{12}-\frac{1}{2}(a_8^{13}+a_8^{23})
+(1+2\cos2\gamma_3)a_8^{12}\\
\end{split}\\
\begin{split}
D_3^{12}/g^2=&D_2^{12}
+8g^2\Bigl(\cos(\gamma_1+\gamma_2)+\cos(\gamma_1-\gamma_2)-2\cos2\gamma_3\Bigr)\\
&+6a_2^{12}-6a_3^{12}+(4+2\cos2\gamma_3)(a_7^{12}-a_8^{12})\\
\end{split}
\end{align}
\begin{align}
\begin{split}
D_4^1/g^2=&2g^2\Big(10-\cos4\gamma_2-\cos4\gamma_3+4\cos2(\gamma_2+\gamma_3)
+4\cos2(\gamma_2-\gamma_3)\\
&\hspace{7mm}+2\cos2\gamma_2+2\cos2\gamma_3-10\cos(\gamma_2-\gamma_3)-
10\cos(\gamma_2+\gamma_3)\Bigr)\\
&-4g^2\left(k_{23}+k_{14}\right)-a_2^{13}-a_2^{12}-a_3^{13}-a_3^{12}-3a_4^1+a_5^1\\
&+(2\cos\gamma_2-1)(a_7^{13}+a_8^{13})
+(2\cos\gamma_3-1)(a_7^{12}+a_8^{12})-a_{6}^1\\
\end{split}\\
\begin{split}
D_5^1/g^2=&4g^2\Bigl(8+\cos2\gamma_2+\cos2\gamma_3-3\cos(\gamma_2+\gamma_3)
-3\cos(\gamma_2-\gamma_3)\Bigr)\\
&-8g^2\left(k_{23}+k_{14}\right)+2a_4^1-6a_5^1+2a_{6}^1\\
&-2\Bigl(a_2^{12}+a_2^{13}+a_3^{12}+a_3^{13}+a_7^{12}+a_7^{13}+a_8^{12}+a_8^{13}\Bigr)\\
\end{split}
\end{align}

\bibliographystyle{ieeetr}
\bibliography{gammaref1}

\end{document}